\documentclass[preprint,1p]{elsarticle}
\usepackage{graphicx}% Include figure files
\usepackage{dcolumn}% Align table columns on decimal point
\usepackage{bm}% bold math
\usepackage{hyperref}% add hypertext capabilities

\begin{document}

\title{\bf{Observation of anti-magnetic rotations in $^{104}$Pd}}

\author[sinp]{N. Rather}
\author[tifr]{S. Roy\corref{cor1}}
\ead{santoshroy@gmail.com}
\author[AMC]{P. Datta}
\author[sinp]{S. Chattopadhyay}
\author[sinp]{A. Gowsami}
\author[iit]{S. Nag}
\author[tifr]{R. Palit}
\author[tifr]{S. Pal}
\author[tifr]{S. Saha}
\author[tifr]{J. Sethi}
\author[tifr]{T. Trivedi}
\author[tifr]{H. C. Jain}
\cortext[cor1]{Corresponding author}
\address[sinp]{Saha Institute of Nuclear Physics, 1/AF Bidhannager, Kolkata 700 064, India}
\address[tifr]{Tata Institute of Fundamental Research, Homi Bhabha Road, Mumbai 400 005, India}
\address[AMC]{Ananda Mohan College, 102/1 Raja Rammohan Roy Sarani, Kolkata 700 009, India}
\address[iit]{Indian Institute of Technology, Kharagpur-721302, West Bengal, India}

\date{\today}

\begin{abstract}
The electric quadrupole transition rates for the high spin yrast states of $^{104}$Pd
 have been measured by using the DSAM 
technique. These values decrease with the increase of angular momentum which is a
 signature of anti-magnetic rotation. A numerical calculation based on semi-classical
 particle plus rotor model for anti-magnetic rotation gives a good description of the experimental routhian
 and the transition rates. This is the first observation of AMR in a nucleus other than Cadmium. 
\end{abstract}

\maketitle

Anti-magnetic rotation (AMR) is a novel mechanism for the generation of high angular momentum states in atomic nuclei 
and was first proposed by S. Frauendorf \cite{frau01,frau02}. This excitation mode derives its origin from the shears
mechanism which is responsible for the origin of M1 bands observed in A$\sim$180, 130 and 100 regions
\cite{mas100,clar01}.
In the semi-classical picture of shears mechanism, the total angular momentum is generated by the angular
momenta of the valence protons and neutrons and the single particle configuration is such
that it allows the perpendicular coupling of these two 
angular momentum vectors. Thus, at the band head, the angle between them (shears angle, $\theta$) is $90^\circ$ and
the higher angular momentum states of the band originate from the gradual closing of the two angular momentum vectors
around the total which resembles the closing of a pair of shears. AMR is a special case of symmetric multi-shears
configuration, which is formed by the two angular momentum vectors of a pair of deformation - aligned protons
(neutrons) in time reversed orbits (${\bf j_h}^{(1)}$ and ${\bf j_h}^{(2)}$) and the angular momentum vector of multiple 
rotation - aligned neutrons (protons) ($\bf j_p$). This structure is symmetric because the shears angle ($\theta$)
between ${\bf j_h}^{(1)}$ - ${\bf j_p}$ and ${\bf j_h}^{(2)}$ - $\bf j_p$ is the same. Thus, the higher angular 
momentum states, in case of AMR, are generated by the simultaneous closure of the multi-shear around ($\bf j_p$) and is given by 
\begin{equation}
I_{sh} = {j_p} + 2{j_h}cos\theta
\end{equation}
The symmetry of this shears structure implies that the perpendicular components of the magnetic moment
for the two shears cancel each other. For this reason the dipole transition rate vanishes for AMR. The
cancellation of the magnetic moment has induced the name Anti-magnetic Rotation due to its similarity
with anti-ferromagnetism where the dipole 
moment of one sub-lattice is in opposite direction to the other half leading to the absence of the net magnetic
moment. However, as the $R_z (\pi)$ symmetry is retained,
the rotational structure decays by weak electric quadrupole (E2) transitions. This transition
 rate is given by \cite{clar01}
\begin{equation}
B(E2) = \frac{15}{32\pi}{(eQ)}^2_{eff} {sin}^4\theta
\end{equation}
where $Q_{eff}$ is the effective quadrupole moment of the core (rotor). Thus, the B(E2) rates are expected
to drop with increasing angular momentum for AMR.
\par
This mode was first experimentally observed in $^{106}$Cd by Simons {\sl et al.} \cite{simo01} and 
subsequently in $^{108}$Cd by P. Datta {\sl et al.} \cite{pdat01}. In both the cases the AMR was built
on $\pi (g_{9/2})^{-2} \otimes \nu [(g_{7/2}/d_{5/2})^2 (h_{11/2})^2]$ configuration and the band heads were found to be at $18 \hbar$
and $16 \hbar$ for $^{106}$Cd and $^{108}$Cd, respectively. The AMR mechanism accounted for an angular momentum increase 
of $8 \hbar$ which corresponded to the complete alignment of the two proton holes and the B(E2) rates were found to 
decrease continuously in this spin domain. \par
The interplay between collective and the anti-magnetic rotations was first reported by
Roy {\sl et. al} \cite{sant01} in $^{110}$Cd where the AMR band was built on  $\pi (g_{9/2})^{-2} \otimes \nu (h_{11/2})^2$
configuration. This interplay was found to span over an angular momentum range of $18 \hbar$ of which $10 \hbar$ was 
due to collective rotation.
In this work, a semi-classical particle rotor model calculation could successfully reproduce the observed alignment feature.
The parameters of the model were determined from the observed systematics in $^{106,108}$Cd.
The effect of the interplay was  
concluded from the slower fall of B(E2) rates in $^{110}$Cd as compared to the pure AMR bands in
$^{106,108}$Cd \cite{sant01}. Such interplay was later reported in $^{105,107}$Cd \cite{deep01,deep02}. Apart from Cd isotopes
mentioned above, experimental investigations of $^{100}$Pd \cite{szhu01},
$^{101}$Pd \cite{msug01} and $^{144}$Dy \cite{msug02} have indicated the possibility of
 existence of AMR bands in these nuclei. However, due to the 
absence of lifetime measurements, the bands of these nuclei could not be identified with AMR. The present work
reports  the lifetime measurements of the high spin yrast states of $^{104}$Pd and establishes AMR for the first time in
 a nucleus which is not an isotope of Cadmium.  \par
In order to populate the high spin states of $^{104}$Pd, the $^{13}$C beam of 63 MeV delivered by 14-UD Pelletron at TIFR was
bombarded on 1 mg/cm$^2$ enriched $^{96}$Zr target. The target had a $^{206}$Pb backing of
 9 mg/cm$^2$ thickness. The de-exciting $\gamma$ rays were detected using the Indian National Gamma
Array (INGA) \cite{mura01} which consisted of 18 Compton suppressed clover detectors. Two and higher fold coincidence data 
were recorded by fast digital data acquisition system based on Pixie-16 modules of XIA LLC \cite{rpal01}.
 The corresponding time stamped data were sorted in 
a $\gamma$-$\gamma$-$\gamma$ Cube using the Multi-PArameter time stamped based Coincidence Search 
program (MARCOS), developed at TIFR. The Cube was used to established the low lying levels of $^{104}$Pd exhibiting no 
lineshapes with the help of the RADWARE program LEVIT8R \cite{dcra01}. It was also used to determine
 their relative intensities. An asymmetric
$90^{\circ}$-vs-all $\gamma$-$\gamma$ matrix was constructed by placing the $\gamma$-energy
 detected at $90^{\circ}$ along one axis and the
coincident $\gamma$-energy detected at any other angle along the other axis. The $90^{\circ}$ projections of different
 $\gamma$-gates on this matrix were used to extend the level scheme to higher spins
 and to measure the corresponding relative intensities since
the angle summed symmetric cube was not suitable for the extraction of
 intensities of the $\gamma$ rays with lineshapes. The level
scheme and the extracted intensities from the present work are in
 agreement with the earlier report\cite{sohl01}. The level scheme
of the yrast cascade of $^{104}$Pd is shown in Fig. 1 where the widths of the transitions are proportional 
to their relative intensities. 
\par
The MARCOS was also used to construct two more angle dependant $\gamma$-$\gamma$ asymmetric 
 matrices for the angles ${40}^{\circ}$ and ${157}^{\circ}$.
These two matrices were used to extract the lineshapes for the levels 
above $I=14 \hbar$ of $^{104}$Pd at the forward and the backward angles. The 
theoretical lineshapes were derived using the code LINESHAPE by Wells and Johnson \cite{jcwe01}.
A Monte Carlo simulation of the slowing down process of the recoiling nuclei in the target and the backing was used to
generate the velocity profiles at the three angles of ${40}^{\circ}$, ${90}^{\circ}$ and ${157}^{\circ}$.
These profiles were obtained  at a time interval of 0.001 ps for 5000 histories of energy losses at different depths
of the target and its backing. The stopping power formula of Northcliffe and Schilling \cite{lcno01} with shell correction
was used for calculating these energy losses. The energies and intensities of the  $\gamma$-transitions were
treated as the input parameters to the lineshape fits. The side feeding intensities were fixed
to reproduce the observed intensity pattern along the yrast cascade shown in Fig. 1. The side feeding intensity to each
level has been modelled as a cascade of five transitions with a moment of inertia which is comparable to that of the band of
interest. The quadrupole moments of the side feeding sequence were allowed to vary which when combined with the moment of
 inertia gave an effective side feeding time for each level. For each observed lineshape, in-band and side-feeding lifetimes
 and the intensities of the contaminant peaks (if present), were allowed to vary. For each set of parameters, the simulated 
lineshapes were fitted to the experimental spectrum using ${\chi}^2$-minimization routine of MINUIT \cite{fjam01}.\par
The topmost 
transition of 1468 keV (${26}^+$$\rightarrow$${24}^+$) was assumed to have 100\% sidefeed. The other parameters
 were allowed to vary until the minimum ${\chi}^2$ value was reached. This led to the estimation 
of the effective lifetime for the ${26}^+$ level. For the ${24}^+$ level decaying through 1365 keV 
transition, the effective lifetime of the ${26}^+$ level and the side-feeding lifetime were considered
 as the input parameters. In this way, each lower
level was added one by one and fitted until all of the observed lineshapes of the yrast cascade of $^{104}$Pd were included
into a global fit where only the in-band and the side-feeding lifetimes were allowed to vary. The uncertainties in the 
measurement were derived from the behaviour of the ${\chi}^2$ fit in the vicinity of the minimum. This procedure of global fit
was repeated at forward (${40}^{\circ}$) and backward (${157}^{\circ}$) angles. Thus, the final value for the level
 lifetime was obtained by taking average from the fits at the two angles. The corresponding 
uncertainty has been calculated as the average of the uncertainties for the two independent
 lifetime measurements for that level added in quadrature.\par
In the present analysis all the lineshapes were extracted from the 803 keV $\gamma$-gated spectrum. In this gate the lineshape
of 927 keV transition (${16}^+$$\rightarrow$${14}^+$) 
was strongly contaminated by 926 keV transition (${6}^+$$\rightarrow$${4}^+$). However, it was found 
from the data that the $\gamma$-transitions below ${10}^+$ state did not exhibit lineshapes. Thus, the
926 keV transition could be treated as a contaminant peak. The intensity of this stopped peak was estimated from the efficiency
corrected areas of 556 (${2}^+$$\rightarrow$${0}^+$), 768 (${4}^+$$\rightarrow$${2}^+$) and 971 (${8}^+$$\rightarrow$${6}^+$) keV
transitions in the 803 keV $\gamma$-gate. These areas were found to be equal within $\pm$1\%. So a stopped peak at 926 keV with
this averaged area was used to determine the true lineshape of 927 keV transition.
The examples of the lineshape fits for the three top transitions in $^{104}$Pd are shown in Fig. 2.\par
The B(E2) transition rates were extracted from the measured level lifetimes using the formula \cite{brow01}
\begin{equation}
B(E2) = \frac{0.0816}{E_{\gamma}^5 \times{\tau}} 
%\label{eqn:eqn2}
\end{equation}
where $E_{\gamma}$ is the energy in MeV of a pure E2 transition, $\tau$ is the level lifetime in pico-seconds and B(E2) is in the units of 
${(eb)}^2$. The extracted B(E2) values are tabulated in Table I where the error bars on the values include the uncertainties on lifetime 
and intensity measurements added in quadrature. The B(E2) values have been plotted as a function of angular momentum 
in Fig. $3(a)$ which show a monotonically falling behaviour. In addition, the values for the ${\mathcal{J}}^{(2)}$/B(E2) ratio for levels 
$I^{\pi}$ = ${16}^+$ and above were found to be an order of magnitude larger than those for a well-deformed collective rotor
and increased with spin. This is expected for an AMR band as the B(E2) values are small
and decrease with spin while the ${\mathcal{J}}^{(2)}$
remains nearly constant. The values of this ratio are also tabulated in Table I. Thus, the high spin levels
of the yrast cascade of $^{104}$Pd seem to  originate due to anti-magnetic rotation.
\par
In order to explore this possibility further, a numerical calculation
based on the framework of semi-classical particle rotor model has been performed. This model has been successfully
employed to describe the observed spectroscopic features of the AMR bands in Cd-isotopes \cite{sant01,sant02}. 
In this model, the energy $E(I)$ is given by :
\begin{equation}
E(I)=\ \frac{(\bf{I-j_{\pi} - j_{\nu}})^2}{2\Im}+\ \frac{V_{\pi \nu}}{2}(\frac{3 {\cos^2 \theta}-1}{2})\ 
+\  \frac{V_{\pi \nu}}{2}(\frac{3 {\cos^2 (-\theta)}-1}{2}) - \ \frac{V_{\pi \pi}}{n} (\frac{3 {\cos^2 (2\theta)}-3}{2})
\end{equation}
where the first term represents the rotational contribution and $\Im$ is the associated moment of inertia. The second and 
the third terms signify the repulsive interaction between the neutron particles and the proton holes and $V_{\pi \nu}$ 
is the interaction strength. The fourth term is the proton - proton (hole - hole) attractive interaction and has been
assumed to be of the same form with the additional boundary condition that it vanishes for $\theta$ = $0^{\circ}$. 
This condition also implies that the attractive particle - particle interaction is absent. There is a scaling factor
$n$ between $V_{\pi \nu}$ and $V_{\pi \pi}$ and is determined by the actual number of particle - hole pairs 
for a single particle configuration. The systematic study of AMR in
 even - even Cd isotopes has indicated that the strength of the 
particle - hole and hole - hole interactions are 1.2 and 0.15 - 0.2 MeV, respectively
 \cite{sant01}. Since $^{104}$Pd is the immediate
even - even neighbour of $^{106}$Cd, it is therefore expected that the same interaction strength should be valid.\par
The angular momentum generated by the interplay between collective rotation and AMR can 
be calculated by imposing the energy minimization condition as a function of $\theta$ on Eq. (3), which gives:
 \begin{equation}
I=\ aj+\ 2j \cos\theta+\frac{1.5 \Im V_{\pi\nu} \cos \theta}{j}-\frac{6 \Im V_{\pi \pi} \cos 2\theta\  \cos\theta}{nj}
%\label{eqn:eqn2}
\end{equation}   
or,
 \begin{equation}
I= I_{sh} + \Im \omega_{sh}
%\label{eqn:eqn2}
\end{equation}  
where, $j$ = $j_{\pi}$, $a$ = $j_{\nu}$/$j_{\pi}$, $I_{sh}$ is the sum of the first two terms of 
Eq. (5) and $\omega_{sh}$ = $\left(\frac{dE_{sh}}{d\theta}\right)
/\left(\frac{dI_{sh}}{d\theta}\right)$ which represents the frequency associated with 
the shears mechanism and is given by
\begin{equation}
\omega_{sh}=\ (1.5 V_{\pi\nu}/j)\cos\theta -(6V_{\pi\pi}/nj)\cos2\theta \ \cos\theta
\end{equation}
Thus, in this model the total angular momentum is generated by the shears mechanism ($I_{sh}$) and the interplay 
between shears mechanism and collective rotation represented by $\Im \omega_{sh}$ in Eq. (6).
 The magnitude of  $\Im$ can be estimated from the equation
\begin{equation}
\Im \omega_{sh}|_{(\theta=0^\circ)}=I_{max}-{I_{sh}}^{max}
\end{equation} 
where $I_{max}$ is the highest observed angular momentum state and $I_{sh}^{max}$ is the maximum angular momentum generated
by the full closure of the double shear. The rotational
frequency $\omega$ is given by
\begin{equation}
\omega = \omega_{rot} - \omega_{sh}
\end{equation} 
where $\omega_{rot}=\frac{1}{2\Im_{rot}}(2I+1)$ is the core rotational frequency
and $\Im_{rot}$ is the core moment of inertia, whose value can be estimated from the slope
of the $I(\omega)$ plot for the ground state band (before the neutron alignment).
Thus, the parameters for the model for $^{104}$Pd can be fixed from the experimental data or the systematics
of the mass region.\par
In the present calculation the configuration for $^{104}$Pd was assumed to be
$\pi g_{9/2}^{-2} \otimes \nu [h_{11/2}^2, ({g_{7/2}/d_{5/2})}^2$] which is the same as for the AMR band of 
its isotone, namely $^{106}$Cd.
For this configuration, the symmetric shear is formed between ${j_h}^{(1)}$ = ${j_h}^{(2)}$  = $j_{\pi}$ = 9/2 and
$j_p$ = $j_{\nu}$ = 16. Since there are eight possible particle - hole and one hole - hole pairs, $n$ = 8. Thus, the
shear parameters from the calculation of the shear angle and I$(\omega)$  plot were $j$ = 4.5$\hbar$, $a$ = 3.55,
 $V_{\pi \nu}$ = 1.2 MeV and $V_{\pi \pi}$ = 0.2 MeV. For $^{104}$Pd, $I_{sh}^{max}$ = $j_\nu$ + 9/2 + 7/2 = 24$\hbar$ and from
experiment $I^{max}$=26$\hbar$, which implied that $\Im \omega_{sh}|_{(\theta=0^\circ)}$ = 2$\hbar$. 
This led to $\Im$ = 5 ${MeV}^{-1}{\hbar}^2$. ${\Im}_{rot}$ has been found to be 17${MeV}^{-1}{\hbar}^2$ from  the 
slope of the ground state band. With these fixed set of parameters, the numerical values of I($\omega$) were calculated 
and have been shown by the solid line in Fig. 3(b) while the experimental values are shown as filled squares.
The calculated values were shifted by the experimental band - head 
frequency of 0.46 MeV. The comparison plot shows that the numerical values obtained from semi-classical particle
rotor model are in good agreement with the experimental values. It is worth noting that in the present model the shears
angle $(\theta)$ is the only variable and every angular momentum state corresponds to a unique $\theta$. At the band
head $\theta$ = ${90}^{\circ}$, I = $j_\nu$ = 16 $\hbar$ (from Eq. 5) and ${\omega}_{sh}$ = 0 (from Eq. 7). The high 
angular momentum states are generated by closing the shears angle and this process has been depicted pictorially 
in Fig. 4 for $^{104}$Pd. The figure shows the shears configuration for the levels of the AMR band which 
have been characterized by their energy and angular momentum.\par
The B(E2) values were calculated using Eq. (1) where $eQ_{eff}$ for $^{104}$Pd was found to be
1.3 $eb$ from the single particle quadrupole moments of the Nilsson states with 
Woods-Saxon potential at $\beta_2$ = 0.19 \cite{sohl01}.
This prescription has been described in detail in ref. \cite{pdat01}. The shears angle for each state
was determined from Eq. (5) and has been given in Fig. 4. The experimental and calculated values have been shown in Fig. 3(a)
and the good agreement provided the essential consistency check for the numerical calculations. It is worth noting
that though these calculations involved a number of parameters but none of them was left free to 
obtain the good agreement between the experimental and calculated values. 
\par
In summery, the lifetimes of the high spin levels of the yrast cascade of $^{104}$Pd have been measured using
 the DSAM technique. The measured quadrupole transition rates exhibit a monotonically decreasing behaviour
with increasing angular momentum in the domain of 16$\hbar$$\leq$I$\leq$26$\hbar$. In addition, the  observed 
${\mathcal{J}}^{(2)}$/B(E2) values for these levels are large. 
These feature are definitive indications of AMR. The numerical calculations for AMR based on semi-classical particle
rotor model provides a good description of the experimental I($\omega$) and B(E2) values. In these calculations, the strengths of
particle - hole and hole - hole interactions in $^{104}$Pd were assumed to be the same as were found for the Cd
isotopes. Thus, AMR has been observed in a nucleus other than Cadmium for the first time which establishes AMR as an 
alternate mechanism for the generation of high angular momentum states in atomic nuclei.
\par  
Authors would like to thank the technical staff of TIFR-BARC pelletron facility for its smooth operation 
throughout the experiment. The help and cooperation of members of the INGA collaboration for setting up 
the array are acknowledged. This work was partially funded by the Department of Science and Technology, 
Government of India (No. IR/S2/PF-03/2003-III). N. R. and P. D. (grant no. PSW-26/11-12) would also like 
to thank UGC for research support.

\newpage
\begin{figure}[h]
\centering
\includegraphics[scale=0.65, angle=0]{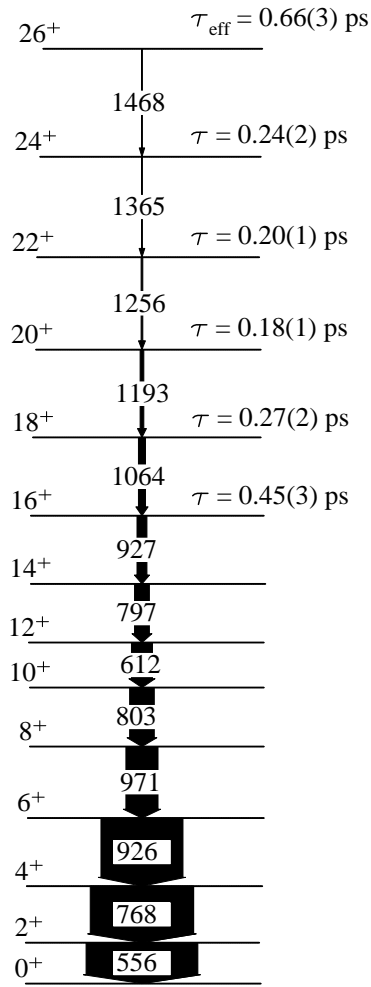}
\caption{The level structure of the yrast cascade of $^{104}$Pd. The energies of the transitions are given in keV and the 
width of the arrow is proportional to the relative $\gamma$-ray intensity.} 
\end{figure}

\newpage
\begin{figure}[h]
\includegraphics[scale=0.6, angle=0]{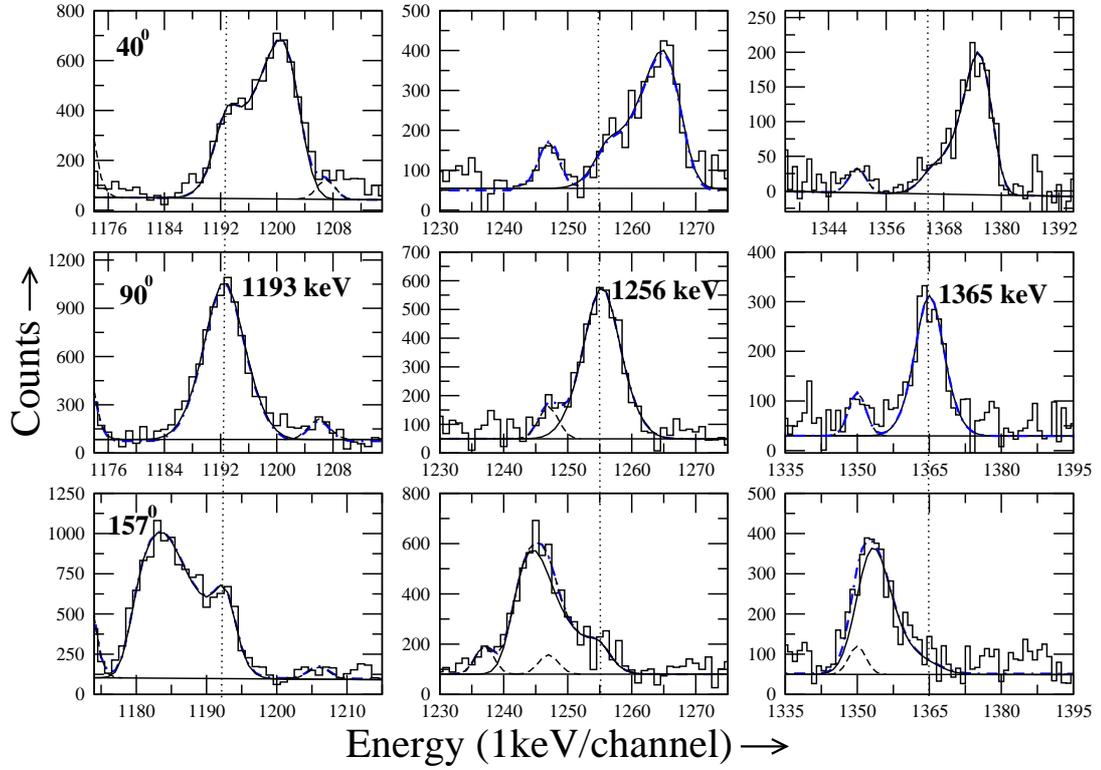}
\caption{Examples of the lineshape fits for 1193 (${20}^+\rightarrow{18}^+$) keV, 1256 (${22}^+\rightarrow{20}^+$) keV and 
1365 (${24}^+\rightarrow{22}^+$) keV transitions  at ${40}^{\circ}$, ${90}^{\circ}$ and ${157}^{\circ}$ with respect 
to the beam direction. The fitted Doppler broadened lineshapes are drawn in solid lines while the contaminant 
peaks are shown in dashed lines.The result of the fit to the experimental data is shown in dot-dashed lines.}
\end{figure}

\newpage
\begin{figure}[h]
\centering
\includegraphics[scale=0.6, angle=0]{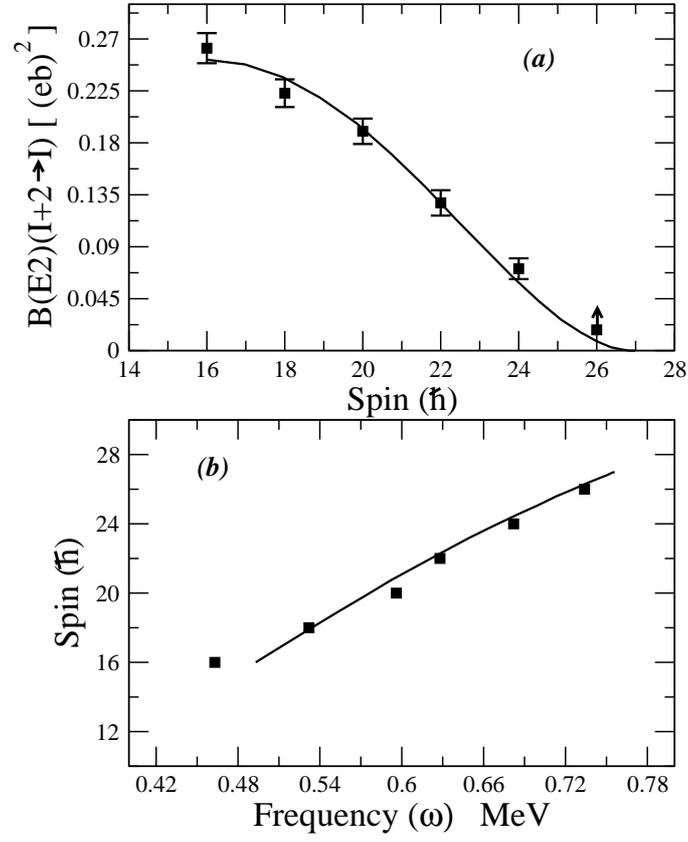}
\caption{The observed B(E2) rates $(a)$ and the I($\omega$) plot $(b)$ in $^{104}$Pd. The lines represent the
numerical values obtained from the semi-classical particle rotor model for the parameter set as given in the text.} 
\end{figure} 

\newpage
\begin{figure}[h]
\centering
\includegraphics[scale=0.6, angle=0]{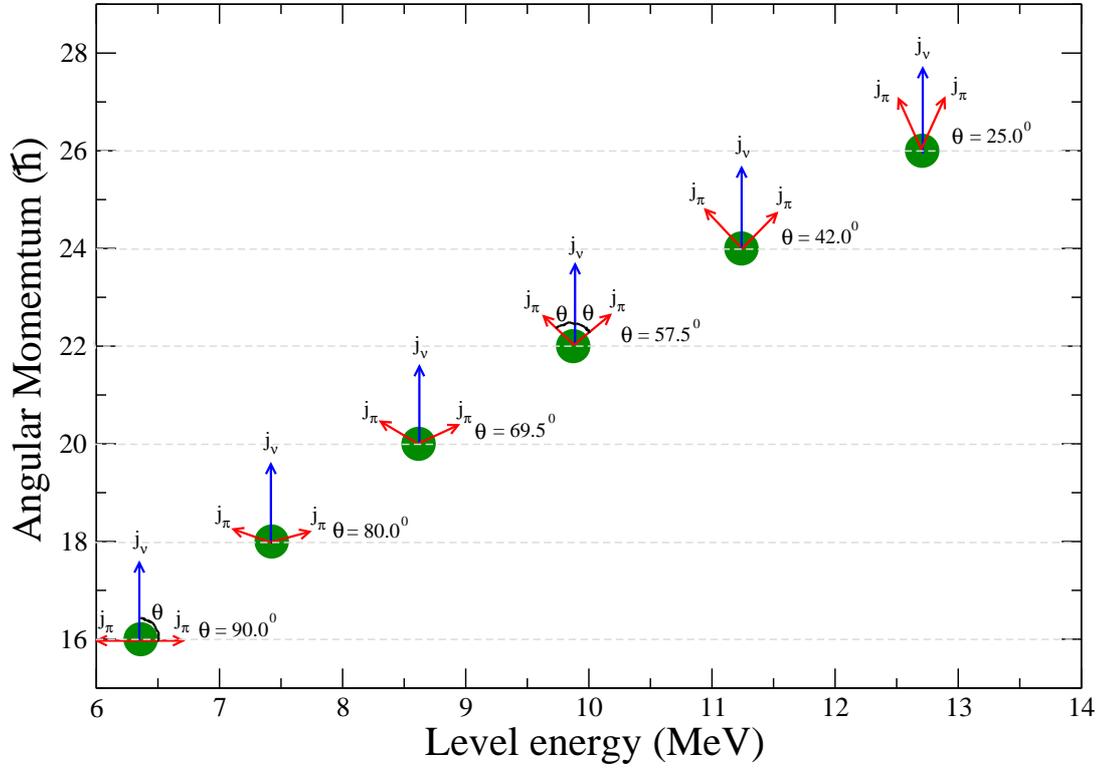}
\caption{Pictorial representation of symmetric shears structure in $^{104}$Pd. The higher spin states are generated
by the gradual closing of the shears angle as depicted in the figure.} 
\end{figure} 
\newpage
\begin{table}[h]
\centering
\begin{tabular}{cccccccc} \hline

Spin &&  Lifetime &&  B(E2) &&${\mathcal{J}}^{(2)}$/B(E2)\\ 
($\hbar$) && ($ps$)       && ${(eb)}^2$&& [${\hbar}^2$ ${MeV}^{-1}{(eb)}^{-2}$]\\ \hline
 ${16}^+$ &&  0.45 (3)  &&  0.26 (2) && 112.3 (8.6) \\ 
 ${18}^+$ &&  0.27 (2)  &&  0.22 (2) && 140.9 (12.8)\\  
 ${20}^+$ &&  0.18 (1)  &&  0.19 (1) && 334.2 (17.6)\\ 
 ${22}^+$ &&  0.20 (1)  &&  0.13 (1) && 282.3 (21.7)\\ 
 ${24}^+$ &&  0.24 (2)  &&  0.07(1)  && 554.3 (79.2)\\ 
 ${26}^+$ &&  0.66 (3)  &&  0.02 (1) \\  \hline
\end{tabular}
\caption{The measured lifetimes and the corresponding B(E2) transition rates for high spin yrast cascade
 of $^{104}$Pd. The last column shows the values of ${\mathcal{J}}^{(2)}$/B(E2) ratio.}
\label{table:table}
\end{table}

\end{document}